\documentclass[%
reprint,
superscriptaddress,
 amsmath,amssymb,
 aps,
 prx,
floatfix,
]{revtex4-2}

\usepackage[english]{babel}

\usepackage[]{hyperref}
\usepackage{graphicx}
\usepackage{dcolumn}
\usepackage{bm}

\usepackage{comment}
\usepackage{physics}
\usepackage{cleveref}
\usepackage{xfrac}

\usepackage{xcolor}
\usepackage{todonotes}
\usepackage{duckuments}
\usepackage{xcolor}
\usepackage{makecell} 
\usepackage[normalem]{ulem} 

\usepackage[ruled,vlined]{algorithm2e}

\definecolor{joakimcomment}{HTML}{53B6E2}

\begin{document}
\date{\today}

\title{Deployed trusted-node quantum key distribution over 300 km with a multi-core fiber access link}

\author{Martin Clason}
\thanks{These authors have contributed equally to this work.}
\affiliation{Department of Electrical Engineering, Linköping University, 581 83 Linköping, Sweden}

\author{Joakim Argillander}
\thanks{These authors have contributed equally to this work.}
\affiliation{Department of Electrical Engineering, Linköping University, 581 83 Linköping, Sweden}

\author{Didrik Bergström}
\affiliation{Department of Electrical Engineering, Linköping University, 581 83 Linköping, Sweden}

\author{Daniel Spegel-Lexne}
\affiliation{Department of Electrical Engineering, Linköping University, 581 83 Linköping, Sweden}

\author{Giulio Foletto}
\affiliation{Department of Physics, KTH Royal Institute of Technology, 106 91 Stockholm, Sweden}

\author{Ashraf El Hassan}
\affiliation{Department of Physics, Stockholm University, 106 91 Stockholm, Sweden}

\author{Mohamed Bourennane}
\affiliation{Department of Physics, Stockholm University, 106 91 Stockholm, Sweden}

\author{Onur Günlü}
\affiliation{Lehrstuhl für Nachrichtentechnik, Technische Universität Dortmund, 44227 Dortmund, Germany}
\affiliation{Department of Electrical Engineering, Linköping University, 581 83 Linköping, Sweden}

\author{Katia Gallo}
\affiliation{Department of Physics, KTH Royal Institute of Technology, 106 91 Stockholm, Sweden}

\author{Rui Lin}
\affiliation{Department of Electrical Engineering, Chalmers University of Technology, 412 96 Gothenburg, Sweden}

\author{Guilherme B. Xavier}
\email{guilherme.b.xavier@liu.se}
\affiliation{Department of Electrical Engineering, Linköping University, 581 83 Linköping, Sweden}


\begin{abstract}

Quantum key distribution (QKD) is increasingly considered for deployment in realistic communication networks, where long distances, heterogeneous fiber infrastructure, and coexistence with classical traffic present substantial challenges. Here, we demonstrate trusted-node QKD between Linköping University and the Stockholm hub of the Swedish national quantum communication infrastructure over 270 km of deployed single-mode fiber, extended by a 33 km multi-core fiber (MCF) segment emulating a metropolitan access link, for a total distance of 303 km. The two sub-links use commercial QKD systems whose receivers are interfaced with external superconducting nanowire single-photon detectors, enabling operation at losses beyond those supported by standard internal gated-mode detectors. We operate the link while actively switching the QKD channel between two MCF cores, with co-propagating Ethernet traffic and injected broadband optical noise in the other cores. The results demonstrate the integration of commercial QKD into demanding, dynamically reconfigurable fiber infrastructure relevant to future hybrid quantum-classical networks. Finally, using the generated secret keys, we illustrate how limited and time-varying QKD throughput affects one-time-pad-protected image transmission: image fidelity depends strongly on the available QKD-generated key budget and the choice of compression algorithm, highlighting application-level challenges for QKD-based encryption in realistic scenarios.

\end{abstract}

\maketitle

\section*{Introduction}
Quantum key distribution (QKD) is a technology that allows two distant parties to establish a secret key with information-theoretic security, guaranteed by the laws of quantum mechanics \cite{bennett_quantum_2014,ekert_quantum_1991}. Compared to classical key distribution methods, QKD offers the advantage of being secure against any computational attack, including those from quantum computers.

Traditional cryptographic methods, such as RSA \cite{rivest_method_1978} and ECC \cite{miller_use_1986, koblitz_elliptic_1987}, are based on assumptions on the computational complexity of certain mathematical problems, which, however, are expected to be efficiently solved by quantum computers using Shor's algorithm \cite{shor_algorithms_1994}. Modern schemes, such as post-quantum cryptography (PQC) \cite{bernstein_post-quantum_2017}, are also being developed to resist this kind of attacks, but they still require computational assumptions and may be vulnerable to future advances in algorithms or hardware. QKD can offer a solution to this problem, providing a method for secure key distribution that is not reliant on such assumptions. In contrast to classical cryptography, QKD protocols such as BB84 \cite{bennett_quantum_2014} and E91 \cite{ekert_quantum_1991} utilize the principles of quantum mechanics to ensure that any eavesdropping attempt will inevitably introduce detectable disturbances in the quantum states being transmitted. This allows the communicating parties to detect the presence of an eavesdropper and discard any compromised keys, ensuring that the final shared key remains secure. This physics-backed security model is expected to be future-proof, making QKD a promising solution for secure communication even in the era of quantum computing. Hybrid approaches, combining QKD with PQC, are also being explored to provide layered security and enhance the overall robustness of communication systems \cite{garms_experimental_2024}.

QKD has seen significant advances in recent years, with both commercial and research systems pushing the boundaries of distance, speed, and security \cite{tang_measurement-device-independent_2016, lucamarini_overcoming_2018, boaron_secure_2018,  pirandola_advances_2020, XuRMP2020, Alia2022DVQKDCoexistence, Zheng2026}. It has been successfully demonstrated in various scenarios, including long-distance fiber links \cite{stucki_high_2009, shimizu_performance_2014, Yin2016MDI, frohlich_long-distance_2017, zhang_long-distance_2020, LiU2023TF}, free-space links \cite{vallone_free-space_2014, liao_long-distance_2017, cai_free-space_2024}, and even satellite-based systems \cite{wang_direct_2013,  vallone_experimental_2015, yin_satellite--ground_2017,liao_satellite--ground_2017, liao_satellite-relayed_2018, li_microsatellite-based_2025}. However, the integration of QKD into existing telecommunications infrastructure remains a central challenge \cite{Dynes2019Cambridge, Pittaluga2025LongDistance, Hajomer2025Coexistence, li_integration_2025}. One promising approach for coexisting quantum and classical data streams is the use of spatial division multiplexing (SDM) to increase channel capacity and also enable high-dimensional quantum communication \cite{xavier_quantum_2020}. A particular implementation is based on multi-core fibers (MCFs), which can support multiple spatial channels within a single fiber strand, allowing for increased data capacity \cite{richardson_space-division_2013}. Previous works have demonstrated the feasibility of QKD over MCFs \cite{Dynes2016, Canas2017, Lin2020MCF, yu_routing_2021, Argillander_2025_ECOC}, and recently also in the MCF network in L'Aquila \cite{Zahidy2024, wu_integration_2025}.

However, many of these advances have been demonstrated in controlled laboratory environments, which may not accurately reflect the challenges faced in real-world deployments. In particular, the integration of QKD systems with existing telecommunications infrastructure, such as deployed fibers or MCFs, presents unique challenges that need to be addressed. In this work, we present the demonstration of long-distance QKD deployed in real-world conditions, interfacing with a multi-core fiber spool access link. We employ two commercial QKD systems connected through an intermediate trusted node. The systems have been customized to operate with external superconducting nanowire single-photon detectors (SNSPDs) to cover the long distances of the two sub-links, including the access link. Our setup further demonstrates active switching of the QKD link through two cores of the MCF, demonstrating QKD operation with active core switching and autonomous reinitialization in a space-division-multiplexed access segment, ultimately showing the feasibility of deploying commercial QKD systems in highly demanding and dynamic environments.

\section*{Results}
\begin{figure*}[ht]
    \centering
    \includegraphics[width=\linewidth]{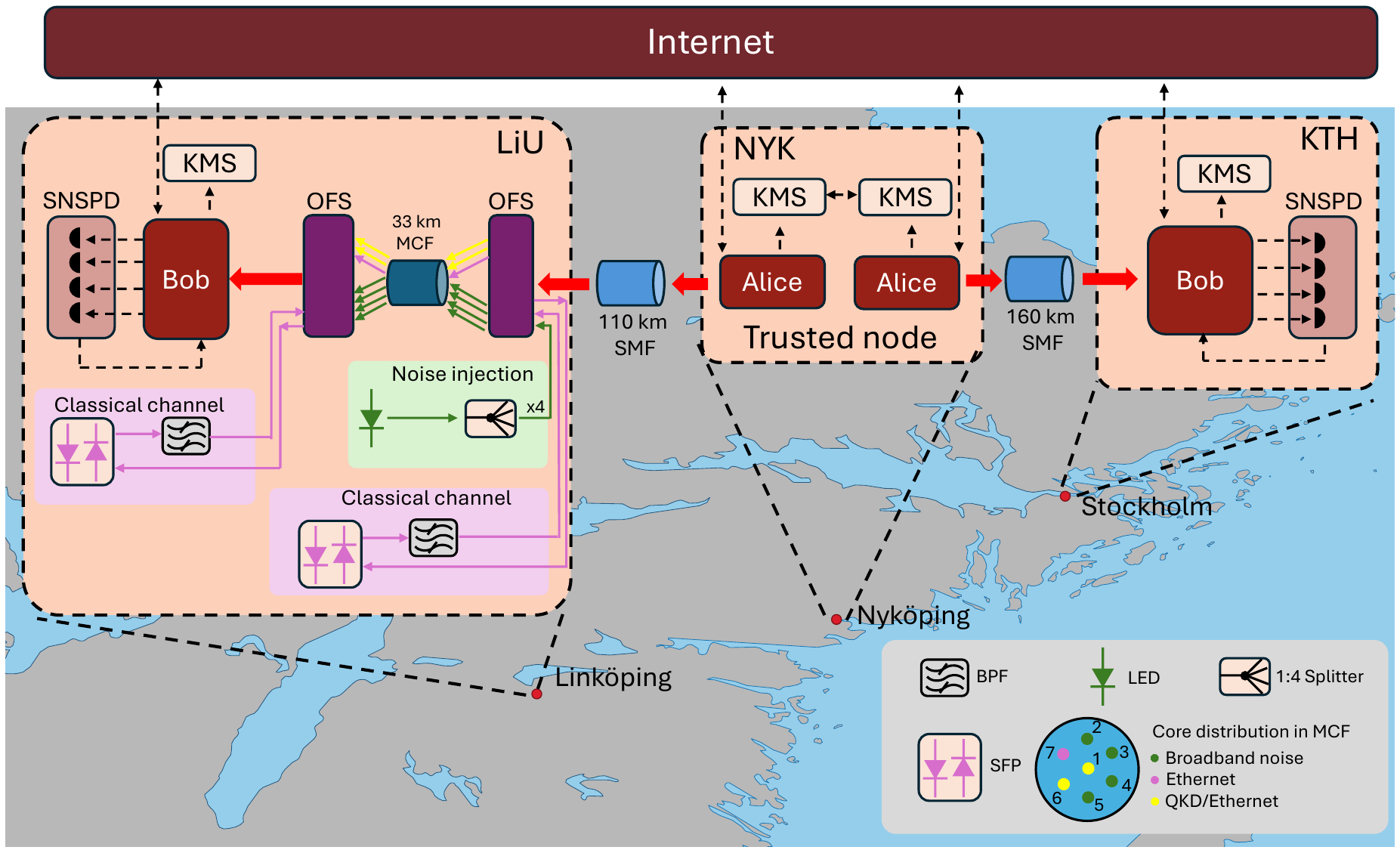}
    \caption{Overview of the deployed experiment in southeastern Sweden. We deploy a long-distance QKD link between Linköping University (LiU) and the the NQCIS Albanova hub at KTH in Stockholm. The two end nodes are connected to an intermediate trusted node located outside of Nyköping (NYK) through deployed single-mode fibers (SMFs). Within the LiU node an additional 33 km multi-core fiber (MCF) spool is employed extending the NYK-LiU link, while allowing spatial multiplexing of classical data channels and additional broadband noise simulating a real populated link.  Both QKD receivers (Bob) have been modified to support external superconducting nanowire single-photon detection (SNSPD) systems. Keys are stored at key management systems (KMSs), located internally within each Alice and Bob unit. The trusted node key exchange protocol is handled automatically through the KMSs, with communication handled through a virtual private network (VPN), deployed over the internet between the sites. Postprocessing information between the QKD units is also done through the same VPN. Please see the text for details. BPF: Band-pass filter; LED: Light emitting diode; KMS: Key management system; OFS: Optical fiber switch; SFP: Small form-factor pluggable optical transceiver. The geographic map background was rendered in QGIS using public-domain vector map data from Natural Earth.}
    \label{fig:map}
\end{figure*}

The deployed QKD link (Fig.~\ref{fig:map}) consists of two end-locations: Linköping University (LiU), and the the Albanova hub of the National Quantum Communication Infrastructure in Sweden (NQCIS) located at the Royal Institute of Technology (KTH) in Stockholm. They are connected through a 270 km long single-mode dark fiber (SMF) link rented from GlobalConnect. The route contains no aerial fiber segments and is divided into two sub-links, with an intermediate trusted node located in a telecommunications exchange node outside of Nyköping (NYK). The first sub-link connects LiU to NYK, covering 110 km with a loss of 23 dB, while the second segment connects NYK to KTH, covering 160 km with a loss of 36 dB. In the LiU node, we employ a 33 km long spooled seven core multi-core fiber (MCF) which simulates a shorter-distance link with co-existing telecommunication traffic, such as an access link. The MCF employed was manufactured at Huazhong University of Science and Technology, with the cores arranged as in the inset of Fig.~\ref{fig:map}. The MCF features a core diameter of 8.7 $\mu$m and a core pitch of 41.1 $\mu$m. Each core is surrounded by a trench with a diameter of 13.5 $\mu$m. The refractive indices of the core, cladding, and trench are 1.4639, 1.4591, and 1.4496, respectively. 

 The cores are individually accessible with fan-in/fan-out (FI/FO) devices. We characterized the fiber distance using an optical time-domain reflectometer (OTDR), and the attenuation of each core, including the fan-in/fan outs as well as the multi-port optical switch (see further details below) using a 1550 nm telecom diode laser and a power meter. The lowest losses were measured for core 1 and core 6 and are 9.3 and 10.5 dB respectively, with the other cores ranging between 11.5 and 13.1 dB. The FI/FO devices and the multi-port switch account for 0.75 and 1.0 dB insertion loss per pass respectively. As cores 1 and 6 have the lowest insertion losses, we assign them to be used for the QKD traffic. We then measured the crosstalk from all the other cores onto core 1 and 6 to be -54.5 and -56.4 dB respectively. 

 The MCF is then inserted before the QKD receiver at LiU, together with a 32 $\times$ 32-port optical fiber switch (OFS - Polatis 6000 family), where we divide the ports to use it as two individual 7 $\times$ 7 switches. The switch is used to dynamically change the core placement of the QKD and the injected classical signals (please see further below for details). With the MCF inserted the total loss of the LiU-NYK link becomes approximately 33 dB and the distance 143 km.

For the QKD systems we employ commercial systems from ThinkQuantum (QuKy EDU Pro). The QuKy systems each consist of a transmitter and a receiver, both corresponding to Alice and Bob respectively. The units have been customized by ThinkQuantum to either employ an internal Indium Gallium Arsenide (InGaAs) single-photon avalanche detection module with typical detection efficiencies of around 10\%, or operate with much higher efficiency external superconducting nanowire single-photon detectors (SNSPDs). The SNSPDs (Appendix) are deployed at both end nodes at LiU and KTH together with the receiver units, while the corresponding transmitters are placed in the trusted node in NYK. Each receiver has an internal passband optical filter (omitted for brevity in Fig. \ref{fig:map}), centered at 1550.12 nm with 0.8 nm passband and $\geq30$ dB suppression, to remove broadband noise that may be present in the fiber link. 

We employ key management systems (KMSs) present in the QKD modules, configured to operate in trusted node mode which allows for seamless key relay between the two segments \cite{DARPA}. In this architecture, the intermediate node at Nyk\"oping is assumed to be trusted. The KMSs use the generated keys 
$K_{(\text{LiU}, \, \text{NYK})}$ 
and 
$K_{(\text{NYK}, \, \text{KTH})}$ 
from the physical links to establish a secure key $K_{(\text{LiU}, \, \text{KTH})}$ 
on the virtual link between LiU and KTH. 

\begin{figure}[ht]
    \centering
    \includegraphics[width=\linewidth]{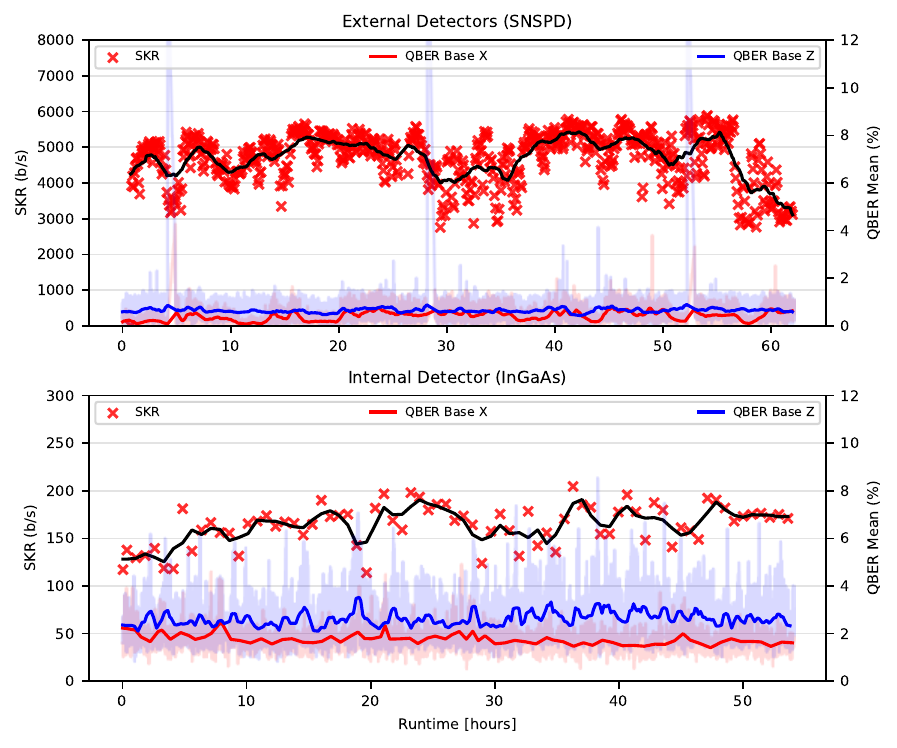}
    \caption{QKD performance between Linköping and Nyköping without the MCF connected. a) Secret key rate (SKR) and quantum bit error rate (QBER) when external superconducting nanowire single-photon detectors (SNSPDs) are connected to Bob. The three clear peaks in the QBER curve correspond to the periodic helium condensation cycle of the SNSPDs at the Linköping node (Appendix), which was set to happen every 24 hours in this measurement. The reported instantaneous values for the QBER are shown in faded blue and red colors, while the average values are otherwise shown as the continuous lines. b) Same results but when using the internal InGaAs avalanche single-photon detector, where we clearly observe the much lower average SKR and slightly higher QBER.  }
    \label{fig:results_qkd_baseline}
\end{figure}

We first demonstrate successful key exchange over the individual sub-link LiU-NYK, without the switch and MCF installed. Furthermore we explicitly show the difference in performance in the link, comparing the use of the internal InGaAs detector with the external SNSPDs (Figs. \ref{fig:results_qkd_baseline}(a) and (b)). We achieve a secret key rate (SKR) of 
$0.16 \pm 0.02$ kbit/s and 
$4.75 \pm 0.71$ kbit/s 
with a quantum bit error rate (QBER) of $2.1 \pm 0.4$ \% and $0.5 \pm 0.2$ \% for the 110 km link for the InGaAs and SNSPD detectors respectively. This value of the QBER in the SNSPD case does not include the peaks observed approximately every 24 h, which originate from the periodic helium evaporation cycle of the SNSPD system deployed at the LiU node (Appendix). As expected, much higher key rates are possible with the SNSPDs for the deployed link, due to the higher detection efficiency, faster recovery time and the fact there is a single internal detector (with time multiplexing), while in the external case four parallel SNSPDs are used. We additionally attribute the improved QBER in the SNSPD case to their lower dark count level.

We then connect the MCF access link to the end point of the deployed link at the LiU node as shown in Fig.~\ref{fig:map}.  The QKD traffic is periodically switched between cores 1 and 6 during the experiment. We inject a classical 10 Gbit/s Ethernet optical channel using dual fiber small-form-factor pluggable (SFP) modules (1546.12 nm with 0 dBm launch power), where one core is employed for the transmitter and the other for the receiver (Fig.~\ref{fig:map}). At each SFP output a dense wavelength division multiplexer (DWDM) is employed as a band-pass filter (BPF), giving $> 50$ dB isolation. The use of this filter, combined with the internal BPF in the QKD receiver and the isolation from the switch and the MCF, was sufficient to ensure no noise from the Ethernet channel disturbed the QKD link, which was running continuously throughout the entire experimental run. 

One of the SFP fiber channels is assigned to core 7, with the other channel alternating between cores 1 and 6, demonstrating dynamic core assignment between QKD and classical signals, opening up further options for hybrid QKD-classical network management systems depending on traffic load. 
We simultaneously create additional crosstalk noise simulating contamination from broadband spontaneous emission generated from other classical traffic that could be present in the other cores of the MCF. This is done by injecting light produced from a light emitting diode (LED) operating at 1550 nm (50 nm bandwidth) into cores 2-5 of the MCF. The LED output is split onto four outputs using three 50/50 fiber couplers in cascade, and these are connected to the cores passing through the switch, generating crosstalk on the QKD core. We changed the driving current to the LED in different steps between 0 and 45 mA corresponding to a maximum optical power of -24.6 dBm, which was the limit of the generated crosstalk that the QKD system could tolerate, given the total experimental loss.  

We then operate the entire QKD link between LiU and KTH over a total of 92 hours, while actively alternating the QKD channel and one of the 10 Gbit/s Ethernet channels between cores 1 and 6 of the MCF, while changing the LED driving current. Due to the extra attenuation induced from the switch and the MCF spool in Linköping, the SNSPDs were used in order to reach a positive SKR. This illustrates the role that SNSPDs can play in deployed QKD receivers, where additional loss margin may be required to accommodate future network extensions, optical switches, or multiplexing components added after the initial deployment. At the KTH site on the other hand, the link could only hold a positive key rate with the SNSPDs. 

For each LED current, starting from 0 mA, we set the QKD system to produce 50 blocks of 50 kbits of raw key material across the entire link. Then we move on to the next current level and so forth until we reach the maximum value of 45 mA. At this point the switch swaps the QKD link from core 1 to core 6, and consequently the Ethernet channel is moved from core 6 to 1. When this happens, the QKD session stops automatically as Bob detects a sudden and quick interruption of the transmission as well as a random polarization rotation on the channel, since the optical path has changed. The system then begins the normal initialization procedure for starting a new QKD session, including resynchronization and basis alignment, which can take between several tens of seconds to minutes, depending on the current photon detection rate.

\begin{figure*}[ht!]
    \centering
    \includegraphics[width=\linewidth]{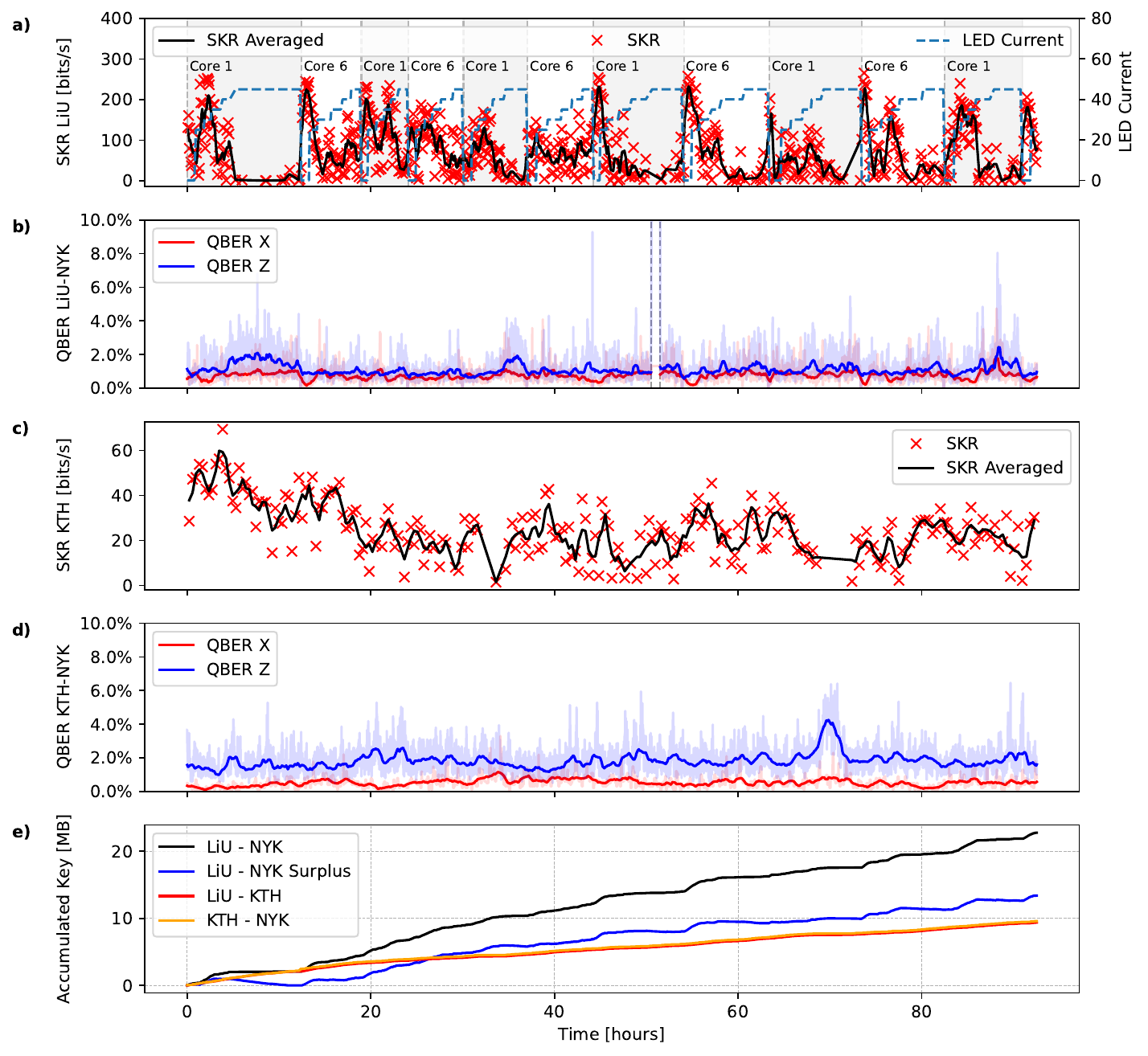}    
    
    \caption{QKD field-trial with active spatial multiplexing in a MCF access link. a) SKR for the Linköping-Nyköping link, with the QKD channel passing through either core 1 or 6 of the MCF, with the other core carrying the downstream channel of a 10 Gbit/s Ethernet optical connection. The upstream channel is fixed at core 7. The other four cores carry broadband noise generated from an LED, whose current we gradually increase within each core cycle, as shown in the figure.  b) Reported quantum bit error rate (QBER) for the two measurement bases ($\mathbf{X}$ and $\mathbf{Z}$). The gap in the measurement around noon on the third day shown explicitly in b) comes from the condensation cycle of the coldhead in the SNSPDs in the LiU node (Appendix). c) and d) SKR and QBER for the Nyköping-Stockholm link over the same time period. The measurement gap is not present here, as the SNSPDs in the Stockholm node operate in continuous mode. e) Accumulated secret key during the measurement run for the individual physical sub-links, LiU–NYK and NYK–KTH, and for the end-to-end LiU–KTH connection established through the trusted-node protocol. The LiU–NYK surplus curve shows the unused key material remaining in the LiU–NYK buffer after the trusted-node key relay has consumed the key needed for the end-to-end LiU–KTH link. This surplus increases when the LiU–NYK sub-link generates key faster than the NYK–KTH sub-link, and is consumed when the LiU–NYK SKR temporarily drops, thereby smoothing short-term fluctuations in the end-to-end key generation. See the text for details.}
    \label{fig:results_liu}
\end{figure*}

We simultaneously monitor the SKR and QBER throughout this measurement, and the results are plotted in Figs.~\ref{fig:results_liu}(a) and (b) for the LiU-NYK sub-link. The same metrics are plotted for the NYK-KTH section in Figs.~\ref{fig:results_liu}(c) and (d). The SKR data points are reported by the system when a successful block of secure key is generated, which fluctuate as the conditions in the channel change. We also show the accumulated secret key material over time for the entire link as well as for each individual sub-link (Fig.~\ref{fig:results_liu}(e)). The QBER is reported directly by the QKD system, and is polled every 4 s. Additionally, Fig.~\ref{fig:results_liu}(a) shows the injected LED current, showing that the SKR goes down as the crosstalk level increases as a function of the LED current as expected, with the switching times between the MCF cores indicated when the current drops from 45 to 0 mA.  At the LiU node, we observe a mean QBER of 
$0.8\% \pm 0.4\%$ and $1.1\% \pm 0.6\%$ 
in the $\mathrm{X}$ and $\mathrm{Z}$ bases respectively and a mean SKR of 
87.2 bit/s $\pm$ 53.4 bit/s. 
At the KTH node on the other hand, a QBER of 
$0.5\% \pm 0.3\%$ and $1.8\% \pm 0.8\%$ 
for  the $\mathrm{X}$ and $\mathrm{Z}$ bases, and a mean SKR of 
24.6 bit/s $\pm$ 10.3 bit/s 
are obtained. At both nodes, we connect the two SNSPDs which have higher efficiency and higher noise to the $\mathrm{Z}$ basis, while in the $\mathrm{X}$ basis, we employ lower noise SNSPDs (Appendix). Furthermore the $\mathrm{Z}$ basis uses two transmitted states, while the $\mathrm{X}$ basis only one, and thus alignment in the former may not be as good as in the latter. These two facts may explain why the $\mathrm{Z}$ basis QBER is consistently higher than the $\mathrm{X}$ one.

We further use the measurement to illustrate how the key buffers in the KMSs respond to the independently fluctuating SKRs of the two physical sub-links. Fig.~\ref{fig:results_liu}(e) shows the accumulated key generated on the LiU–NYK and NYK–KTH sub-links, together with the end-to-end LiU–KTH key established through the trusted-node protocol. Since each bit of end-to-end key requires available key material on both physical sub-links, the trusted-node key rate is ultimately constrained by the sub-link with the lower available key supply. However, because the KMSs include buffers, excess key material generated by the faster sub-link is retained as surplus in its local buffer rather than being discarded.

Over the full measurement run, the LiU–NYK sub-link generates key at a higher average rate than the NYK–KTH sub-link, so the LiU–KTH end-to-end key accumulation rate is usually limited by the NYK–KTH section. This is reflected by the buildup of surplus key in the LiU–NYK buffer. The usefulness of this buffering is most clearly seen during the first day, when the LiU–NYK SKR temporarily stagnates, as also visible in Fig.~\ref{fig:results_liu}(a). During this interval, the end-to-end LiU–KTH key can nevertheless continue to accumulate because previously stored LiU–NYK surplus key is consumed. The LiU–NYK surplus curve therefore directly shows how the buffer absorbs temporary imbalances between the two sub-links. 

\begin{figure}[ht!]
    \centering
   
    \includegraphics[width=\linewidth]{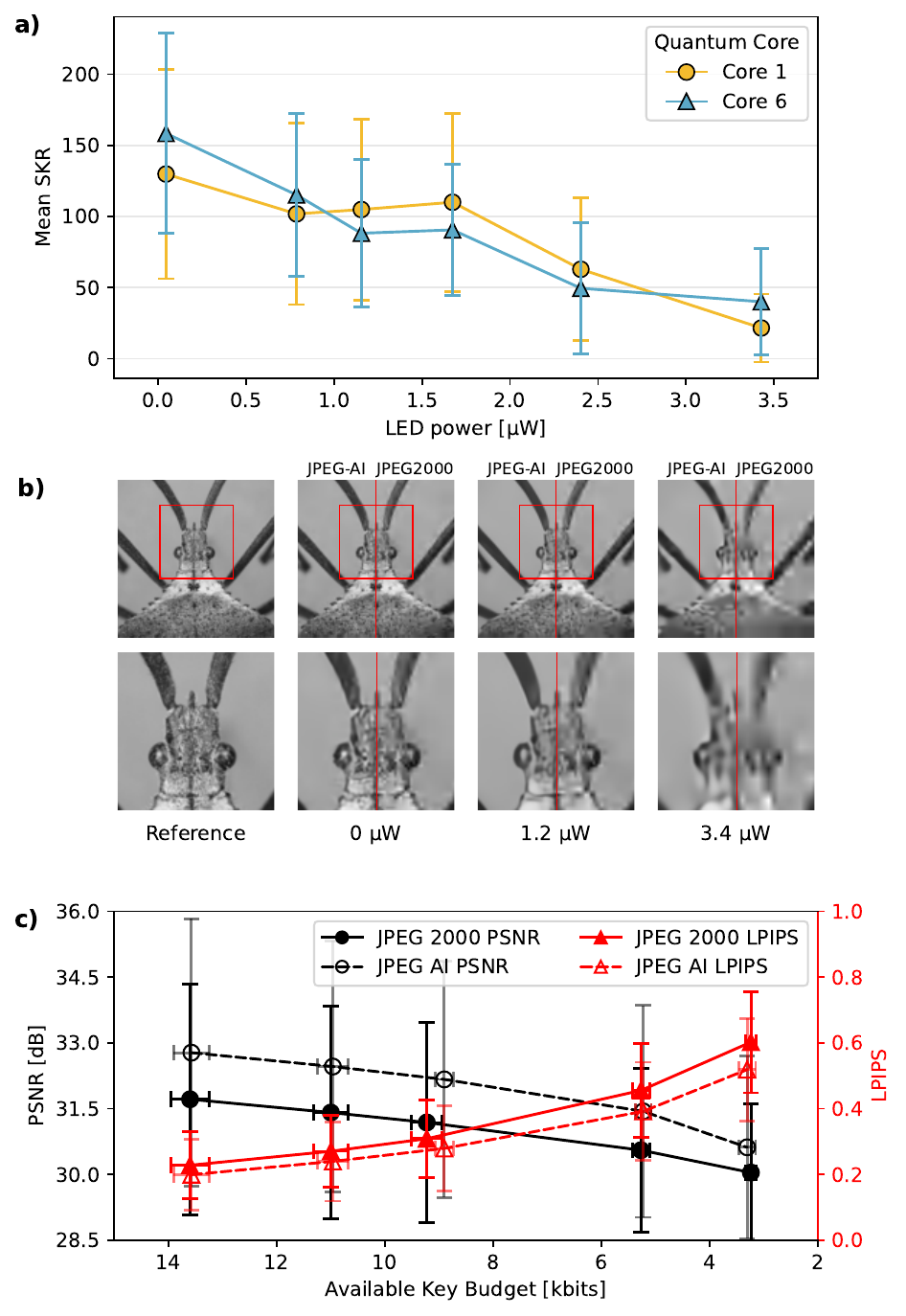}
    
    \caption{SKR performance and information-theoretically secure image transmission. (a) Impact of injected broadband noise from the LED on the deployed long-distance QKD channel connected to the MCF access link. The error bars represent the standard deviation of the SKR distribution for each LED current. (b) We use the available secret key bits to encrypt the compressed reference image using OTP, given the generated key aggregated over a time budget of 100 s. Each image is compressed using both JPEG 2000 and JPEG AI, where we compare the decompression outputs by splitting each image in half and showing the visual results. Adapted from an image by Brett Morgan, University of Texas at Austin, Insects Unlocked project, obtained via Wikimedia Commons; CC0 1.0. (c) Comparison of the average performance between JPEG 2000 and JPEG AI for a dataset of 2100 images in terms of two metrics for image quality, i.e., PSNR and LPIPS. The x-axis indicates the available secret-key budget over a 100 s transmission window, and the vertical error bars show one standard deviation of the corresponding image-quality metric across the 2100 images in the dataset. The horizontal error bars show one standard deviation of the actual compressed image size, which represents the image-dependent variation in the number of actual key bits required for OTP encryption. The available key budget corresponds to the secret key generated under 100 s for the same optical power levels from (a).}
    \label{fig:led_noise_qber_skr}
\end{figure}

In addition to the injected LED noise, there is a fluctuation of background noise photons that we attribute to contamination from traffic present in other fibers in the same cable. The background noise exhibited a fluctuation of $\approx 10$ counts per second over periods of several minutes to hours. This scenario presents typical conditions one may encounter in a real deployed fiber network, where part of the noise may be controllable (or known) by the QKD users (i.e. how much extra traffic is allowed in the same link), while external noise sources coming from other networks that may share the same cable cannot be predicted. We plot the average SKR for each injected current value and for each core over the entire run in Fig.~\ref{fig:led_noise_qber_skr}(a). We nevertheless observe a clear general behaviour where the SKR degrades as a function of the added LED noise. As the QKD link operates close to the maximum attenuation limit with the SNSPDs, there is significant statistical fluctuation in the generated SKR.

As the deployed trusted-node link produces only limited and time-varying SKRs, an application using one-time pad (OTP) encryption must be flexible in terms of the instantaneous SKR. To illustrate this restriction, we evaluate OTP-encrypted image transmission under fixed transmission time windows using two different image compression algorithms. We use a fixed time budget for each image transmission of 100 s for three different average SKRs between the two cores, corresponding to the LED output optical power of 0, 1.2 and 3.4 $\mu$W. Each $256 \times 256$ pixel image is split longitudinally in half (see Fig. \ref{fig:led_noise_qber_skr}(b)), where we visually show the differences between employing the standard JPEG~2000 or the modern JPEG AI compression algorithms (see Appendix). These results show that deep-learning-based image compression algorithms, such as JPEG AI, can significantly improve the efficiency of OTP-encrypted image transmission under highly limited secret-key throughput, a constraint that is especially relevant in real-time QKD-based scenarios but less typical in conventional communication applications. 

We then quantify the average performance of the two compression algorithms by calculating the reconstruction fidelity, using the metric peak signal-to-noise ratio (PSNR), and perceptual similarity, using the metric learned perceptual image patch similarity  (LPIPS) (see Appendix for their definitions). In Fig.~\ref{fig:led_noise_qber_skr}(c), these results are depicted as a function of the available key budget under 100 s by assuming that different mean SKRs are produced by the long-distance QKD link subject to the LED power levels of Fig.~\ref{fig:led_noise_qber_skr}(a). Both metrics are calculated for each key budget level for a 2100 image subset of the NUS-WIDE dataset~\cite{NUSWIDE}, thus showing the expected real-world variation on the performance (see Appendix for more details). In Fig.~\ref{fig:led_noise_qber_skr}(c), as the available secret-key budget decreases, the LPIPS advantage of JPEG~AI over JPEG~2000 becomes larger, indicating that its perceptual-quality gains are more pronounced under stringent QKD secret-key throughputs.

\section*{Discussion and Conclusion}

Our results show that commercial QKD can be operated in a heterogeneous fiber infrastructure that combines long-distance deployed single-mode fiber through a trusted-node key relay, and a dynamically reconfigurable multi-core fiber access segment. The experiment tests QKD under several conditions that are expected to arise in practical future quantum-secured telecommunication networks, including high channel loss, spatial multiplexing, coexistence with classical traffic and broadband noise, as well as reconfiguration of the quantum channel during operation. 

A key feature of the experiment is the use of multi-core fiber as a reconfigurable quantum-classical access link connected to a much longer distance QKD feed link. By actively switching the QKD channel between two cores while a 10 Gbit/s Ethernet channel is simultaneously present in a neighboring core, we emulate a network scenario in which quantum and classical channels are not statically assigned, but instead may be dynamically allocated according to the network management layer \cite{yu_routing_2021}. The observed degradation of the SKR with increasing injected broadband noise shows that such allocation cannot be treated as a purely classical network-management problem. Instead, spatial channel assignment, launch powers, core attenuation, and the instantaneous state of the QKD key buffer must be considered jointly. This points toward the need for a hybrid network control, where QKD parameters are integrated with software-defined networking \cite{Aguado2019SoftwareDefinedQKD}.

The comparison between the internal InGaAs detector configuration and the external SNSPD configuration highlights the importance of detection technology as a practical link-budget parameter. In the tested deployed link, the SNSPDs  provide the additional loss margin needed to operate when including the extra losses from the MCF, fan-in/fan-out devices, and optical switching.  From a deployment perspective, the use of external SNSPDs with commercial QKD systems points to an upgrade path for extending the reach of individual spans, potentially reducing the number of trusted nodes required in a deployed network. At the same time, the observed interruptions associated with the helium evaporation cycle in the detectors at the LiU node, as well as QKD reinitialization resulting from switching operations in the MCF, show that detector availability, autonomous recovery, and polarization re-alignment are relevant parameters at the network management level.

As a practical application of dynamically fluctuating SKRs we illustrate OTP encrypted image transmission under limited and time-varying QKD throughput. For OTP the secret key is consumed at the same rate as the encrypted payload, so the relevant parameter at the network management level is not only whether a positive SKR is achieved but whether sufficient key material is available within the required service time. The comparison between JPEG 2000 and JPEG AI shows that the choice of compression algorithm is important in quantum-secured communication, given that, for a fixed key budget, a more efficient encoding can improve the reconstructed image quality or reduce the waiting time needed to accumulate enough key material. This connects the physical-layer performance of the deployed QKD link to a concrete application metric and suggests that future QKD applications should co-design key generation and buffering, compression, and encryption policies. Overall, our results demonstrate not only long-distance QKD, but the operation of practical applications employing commercial QKD systems in a dynamically reconfigurable, spatial-division-multiplexed optical access link connected to a deployed long-haul trusted-node network.

\section*{Appendix: Materials and Methods}
\subsection*{Image compression algorithms and quality metrics}
We investigate the feasibility of transmitting images that are compressed and then OTP-encrypted, subject to the generated SKR and image-quality constraints. We consider two compression algorithms (codecs): JPEG 2000~\cite{taubman2002jpeg2000} (our baseline), and JPEG AI~\cite{shan2026jpegai}. While the JPEG 2000 standard uses a classical wavelet-based method, JPEG AI applies deep neural networks to enable greater flexibility and improved performance in terms of the required transmission rate and the average distortion in the reconstructed image~\cite{shan2026jpegai}. We evaluate the image quality with respect to two metrics: the peak signal-to-noise ratio (PSNR) and learned perceptual image patch similarity (LPIPS) \cite{zhang2018unreasonable,bergstrom2026ICASSP}, which are next defined.

 Let \(\mathbf{S}, \widehat{\mathbf{S}} \in \mathbb{N}_{\text{P}}^{H\times W} \) be an original and reconstructed grayscale image, respectively, where \(\mathbb{N}_{\text{P}} = \{0,\dots,255\}\) are 8-bit pixel values, and $H$ and $W$ are the image height and width, respectively, which are 256 pixels each. The PSNR metric evaluates the pixel-wise fidelity as \(\text{PSNR}(\mathbf{S}, \widehat{\mathbf{S}}) \triangleq 10\cdot\log_{10}(255^2/\text{MSE}(\mathbf{S}, \widehat{\mathbf{S}}))\), measured in dB and where MSE$(\cdot,\cdot)$ computes the mean squared error. Thus, a smaller MSE corresponds to a smaller average pixel error and, therefore, a higher PSNR. Moreover, the LPIPS metric~\cite{zhang2018unreasonable} is experimentally demonstrated to be an accurate analog to human perception similarity, i.e., the metric agrees with humans on the answer to the question ``which of two distorted images is more similar to the original". The metric ranges from 0 to 1, with lower values indicating higher perceptual similarity. 
It is defined as
\begin{align}
    \text{LPIPS}(\mathbf{S},\widehat{\mathbf{S}})=\sum_l\frac{1}{H_lW_l}\sum_{h,w} \lVert w_l \odot (y^l_{hw}-\widehat{y}^l_{hw}) \rVert_2^2
\end{align}
where \(y^l_{hw},\widehat{y}^l_{hw}\) are the channel-normalized features of the two inputs at spatial location \((h,w)\) in layer \(l\) of the neural-network feature extractor used by LPIPS, \(w_l\) is a learned channel-wise weighting vector, \(H_l\) and \(W_l\) are the height and width of that feature layer and $\odot$ is element wise multiplication respectively.

For each image and compression algorithm, we encode the image subject to the available secret-key budget. Since OTP encryption requires one secret-key bit per payload bit, the compressed payload size directly determines the number of QKD-generated key bits consumed by each image transmission. For a fixed key budget and fixed image dimensions, the realized compressed payload size is not identical across all images. It depends on, e.g., image texture, spatial structure, and codec design. This image-dependent rate variation is a standard feature of transform-based and learning-based image coding \cite{taubman2002jpeg2000, shan2026jpegai}. Therefore, in Fig.~\ref{fig:led_noise_qber_skr}(c), the horizontal error bars indicate the standard deviation of the realized compressed payload sizes over the 2100 NUS-WIDE images. Moreover, the vertical error bars indicate the standard deviation of the corresponding quality metric, i.e., PSNR or LPIPS, over the same image set. In Fig.~\ref{fig:led_noise_qber_skr}(c), we also compare the JPEG~AI and JPEG~2000 performance metrics at approximately equal mean secret-key budgets. These budgets are obtained from the measured mean SKRs under different LED-noise levels over a fixed transmission-time window.

\subsection*{QKD system}

The ThinkQuantum QuKy system implements a 3-state version of the efficient BB84 protocol with decoy states \cite{lo_decoy_2005, wang_beating_2005}, using the polarization degree of freedom. The system operates at a fixed clock rate of 50 MHz with the quantum states generated by an internal distributed feedback (DFB) laser at a wavelength of 1550.12 nm, corresponding to channel 34 in the International Telecommunication Union dense wavelength division multiplexing (ITU-DWDM) grid. The attenuation of the NYK-KTH link is too high for the internal Indium Gallium Arsenide (InGaAs)-based avalanche photodiode single-photon detector used by the QuKy-RX, justifying the use of customized units that can interface with external SNSPD systems. Internally, ThinkQuantum employs an internal passive basis choice at the receiver, based on a 50:50 fiber coupler where the two outputs lead to two polarizing beamsplitters oriented to implement the $\mathrm{X}$ and $\mathrm{Z}$ measurement settings as required in BB84 \cite{Agnesi2020}. Then the four measurement outputs are connected to the external SNSPDs using single-mode fiber optic patchcords, and the electrical outputs from each detector are connected back to the QuKy-RX-unit using coaxial cables. 

The QuKy system also includes internal automatic polarization controllers to compensate for polarization drifts in the fiber link. The polarization controllers are able to undo the polarization transformations induced by the fibre links. The systems are specified to compensate for polarization rotations of at least $0.1$ rad/s, which was sufficient for operation in the employed links.

When operating the QuKy system, the postprocessing block size, was set to $N=50\cdot10^3$ bits, and the target extinction ratio was set to $16 $dB (for both the $\mathbf{X}$ and $\mathbf{Z}$ bases). The postprocessing block size determines the number of raw key bits processed in each round of post-processing, and the results obtained are conditioned on this finite-key size. The extinction ratio is the target extinction of Alice's and Bob's bases respectively, and is used when the units align their respective bases.

In the trusted node the two QuKy-TX units are connected together through a key delivery interface, allowing for the secure transfer of keys within the trusted node between the two segments of the QKD link, which establishes the trusted node key relay. To facilitate the classical communication between the three nodes, we employ a heterogeneous network setup. The LiU and NYK nodes are connected through a layer 2 tunnel over an internet connection (provided to the NYK node by GlobalConnect), effectively creating a virtual private network (VPN) between the two locations. The layer 2 tunnel spans the LiU network and the NYK exchange, allowing for seamless communication between the two nodes. The LiU and KTH nodes are connected through a direct layer 3 tunnel running internet protocol security (IPSec) on Netgate firewalls over the internet. This further allows for the classical communication between the LiU and KTH nodes, which is essential for the operation of the QKD link.  Our setup integrates QKD systems into existing network infrastructure using different tunneling techniques that may be present in real-world deployments. Importantly, we show that routing does not need to be symmetric for the classical channel, as the LiU-NYK and NYK-KTH segments use different tunneling techniques, and that commercial QKD systems are robust against variations in latency and jitter that may arise from the use of internet connections for the classical channel.

\subsection*{External SNSPDs}
\label{appendix:quky_details_snsdp_details}
The external superconducting nanowire single-photon detectors (SNSPDs) used in our setup are critical components for achieving positive SKRs in both of the high-attenuation employed sub-links. The SNSPDs employed at the LiU node are housed within an ID Quantique ID281 system. Two detectors have a high detection efficiency of (92\% and 93\%), with a dark count rate of $\leq30$ counts per second. The other two detectors have ultra-low dark count rates ($\leq 1$ count per second) but a slightly lower detection efficiency of 80\% for both. The SNSPDs are operated at a temperature of 0.8 K using a closed-cycle cryostat. They are sorption-cooled SNSPDs, which utilize a sorption pump to achieve the required low temperatures needed for ultra-low dark count operation. The sorption system operates on an condensation-evaporation cycle of a maximum of 70 hours, after which the liquid helium inside the cryostat needs to be re-condensed, a process which takes approximately one hour. 

The SNSPDs at the KTH node are part of a Single Quantum EOS system, and once again four detectors are employed. Similarly, two detectors have a high detection efficiency of approximately 90\% and a dark count rate of $\leq20$ counts per second. The two other detectors have dark count rates of $\leq 2$ counts per second but a slightly lower detection efficiency of 81\%. The SNSPDs are operated continuously at a temperature of 2.9 K using a closed-cycle cryostat.


\begin{acknowledgments}
We would like to thank Daniel Rovaniemi and Pär Stolpe for technical assistance and Ming Tang for supplying the multi-core fiber. This work is supported by the project “DIGITAL-2022-QCI-02-DEPLOY-NATIONAL” (Project number: 101113375 - NQCIS National Quantum Communication Infrastructure in Sweden) funded by the European Union together with Vinnova and the Wallenberg Centre for Quantum Technology (WACQT). Additional support is provided by Vetenskapsrådet (VR). We also acknowledge helpful discussions with other NQCIS partners in Sweden, other member states' EuroQCI projects and ThinkQuantum. Moreover, the work of O. Günlü was partially supported by the EU COST Action 6G-PHYSEC, Swedish Foundation for Strategic Research (SSF) under Grant ID24-0087, and German Federal Ministry of Research, Technology and Space (BMFTR) 6GEM+ Transfer Hub under Grants 16KIS2412 and 16KISS005.
\end{acknowledgments}

\bibliography{main.bib}

\end{document}